\documentclass[twocolumn]{webofc}
\usepackage[varg]{txfonts}   % Web of Conferences font
\usepackage{amsmath}
\begin{document}
\title{AugerPrime: the Pierre Auger Observatory Upgrade}

\author{\firstname{} \lastname{Antonella Castellina}\inst{1}\fnsep\thanks{\email{antonella.castellina@to.infn.it}} 
        \firstname{} \lastname{for the Pierre Auger Collaboration}\inst{2}\fnsep\thanks{\email{auger_spokespersons@fnal.gov}}
        \fnsep\thanks{Authors list: http://www.auger.org/archive/authors$\_$2018$\_$10.html}}

\institute{ INFN, Sezione di Torino and Osservatorio Astrofisico di Torino (INAF), Torino, Italy 
\and
           Av. San Mart\'in Norte 304, 5613 Malarg\"ue, Argentina}

\abstract {
The world largest exposure to ultra-high energy cosmic rays accumulated by the Pierre Auger Observatory 
led to major advances in our understanding of their properties, but the many unknowns about the nature 
and distribution of the sources, the primary composition and the underlying hadronic interactions prevent 
the emergence of a uniquely consistent picture.
The new perspectives opened by the current results call for an upgrade of the Observatory, whose main aim 
is the collection of new information about the primary mass of the highest energy cosmic rays on a 
shower-by-shower basis. The evaluation of the fraction of light primaries in the region of suppression of the 
flux will open the window to charged particle astronomy, allowing for composition-selected anisotropy searches. 
In addition, the properties of multiparticle production will be studied at energies not covered by man-made 
accelerators and new or unexpected changes of hadronic interactions will be searched for.
After a discussion of the motivations for upgrading the Pierre Auger Observatory, a description of the detector 
upgrade is provided. We then discuss the expected performances and the improved physics sensitivity 
of the upgraded detectors and present the first data collected with the already running Engineering Array.
}
\maketitle
%-------------------------------------------- Introduction
\section{Introduction}
\label{intro}
Although we still lack a comprehensive theory capable of explaining where the ultra-high energy cosmic rays (UHECRs)
are produced, what is the acceleration mechanism driving nuclei to such 
extreme energies and their chemical composition, a large amount of information was gained in the recent years
by exploiting the data taken with the Pierre Auger Observatory \cite{PAO1} . 

The suppression of the cosmic ray flux for $E > 4 \times 10^{19}$ eV has been unambiguously established; 
the current limits on the photon and neutrino fluxes at ultra-high energy strongly limit the possibility that
the observed flux is produced in non standard ``top-down'' processes \cite{unger}. \\
A dipolar large scale anisotropy has been observed in cosmic rays with $E>8$ EeV, 
proving that they are indeed of extragalactic origin and opening a window on the study of the properties of 
extragalactic magnetic fields \cite{dipole}.
Strong hints of correlation with the direction of bright, nearby starburst galaxies
have been obtained in intermediate-scale anisotropy searches, while an excess of events have been seen for $E>5.8 \times 10^{19}$ eV
in the direction of the Centaurus A radio galaxy \cite{SBG}.

In the standard astrophysical Scenario, the UHECRs are mainly extra-galactic protons and the observed suppression of 
the flux is ascribed to the energy losses suffered by the particles during propagation, after having been 
accelerated with a power-law injection spectrum $E^{-2}$ in sources distributed as the matter  
in the Universe (``photo-disintegration Scenario''). 
However, the energy at which the integral flux measured at Auger drops by a factor of two below
what would be expected without suppression, $E_{1/2}$ = (23$\pm$1(stat.)$\pm$4(syst.)) EeV, is quite different
from the  $E_{1/2} = 53$ EeV predicted for this scenario. 
Furthermore, this model is challenged by the results obtained on the primary composition \cite{comp1,comp2,comp3}. 
Indeed, the studies of the evolution with 
energy of the shower development in the atmosphere point to a mass composition becoming progressively
heavier with increasing energy. The heaviest (iron) component appears almost unnecessary, while the protons 
add up to not more than 10\% at the highest energies. 

The best fit to the spectrum and composition as measured at the Pierre Auger Observatory is obtained for a model
in which  the sources accelerate particles to maximum 
energies proportional to their charge, with a hard injection index $\gamma \sim 1$. The observed suppression 
in the flux would then be ascribed to a cutoff in the energy spectra of the sources (``maximum rigidity scenario'') \cite{scenarios}.

The interpretation of the results in terms of composition is made even more difficult by the large uncertainties
in predicting hadronic multiparticle production at energies beyond those accessible in the laboratory. 
Different studies aiming at the evaluation of the muon content in air showers \cite{muon1,muon2,muon3},
exploiting the study of hybrid events \cite{muon4}, or analysing the azimuthal asymmetry in the risetime of signals
in SD  \cite{muon5}, found inconsistencies
in the modeling of hadronic interactions and muon production as implemented in the simulations  
.
%%%% Fig.1
	\begin{figure*}[t]
	\centering
	\includegraphics[width=0.8\linewidth]{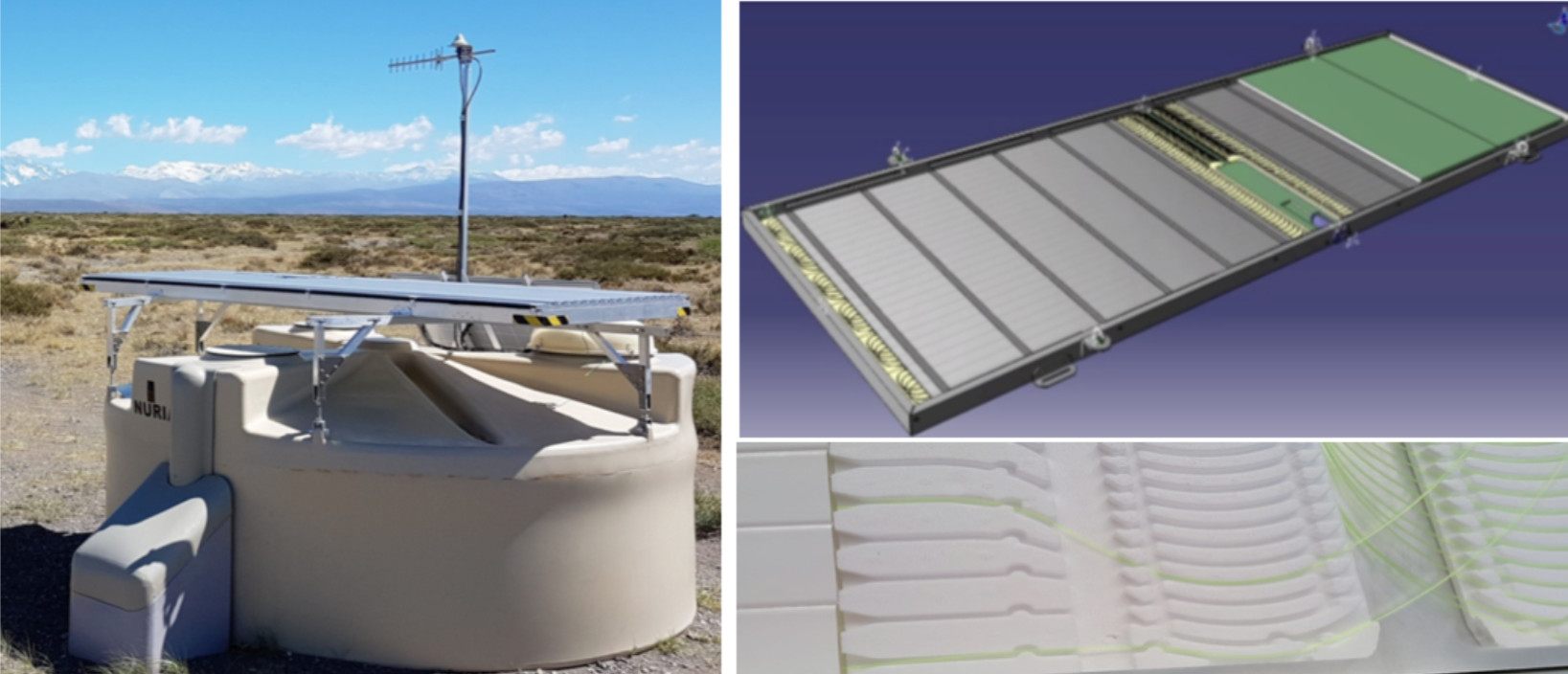}
	\caption{Left: One station of the AugerPrime Engineering Array, featuring the Surface Scintillator Detector 
	(SSD) on top the WCD. Top-right: layout of the SSD. Bottom-right: details of the fibers in the router.}
	\label{fig_1}
	\end{figure*}

An upgrade of the Pierre Auger Observatory, dubbed ``AugerPrime'' \cite{PrimeArch,PrimeGen}, has been planned to 
address the  following open questions :

$\bullet$ understand the origin of the flux suppression, by differentiating between a cut-off due to propagation effects  
and the maximum energy reached in the sources. This will allow us to  provide fundamental constraints on the 
sources of the ultra high energy cosmic rays and their properties. 

$\bullet$ evaluate the possible existence of a fraction of protons at the highest energies. The answer to this fundamental
question will set the stage for future experiments and assess the feasibility of charged particle astronomy.

$\bullet$ add information about hadronic interactions in a range of energy and in a kinematic region which is not
accessible in man-made accelerators. Non-standard mechanisms, such as Lorentz invariance violation or
extra dimensions will also be searched for.

To reach these goals, it is essential to obtain additional measurements of composition-sensitive 
observables on a high statistics sample of ultra-high energy events, thus exploiting the Surface Detector of the
Pierre Auger Observatory.

%-------------------------------------------- Elements of AugerPrime
\section{Elements of AugerPrime}
\label{sec1}

The key elements of the upgrade consist of new plastic scintillator detectors on top of 
the water-Cherenkov stations (WCD) of the surface array (SD),  an additional small
photomultiplier (PMT) installed in the WCD for the extension of the dynamic range, and new SD electronics.\\
Additionally an underground muon detector is going to be deployed aside each WCD in the infill region of the 
Observatory, to provide direct muon measurements.
The  upgrade will also be complemented by extending the measurements of the existing Fluorescence 
Detectors (FD) into periods of higher night-sky background, to increase their duty cycle.
Finally, based on the Radio Engineering Array (AERA) experience and results, a new project for adding
a radio detector  to AugerPrime is now on-going.

%-------------------------------------------------------- SSD
\subsection{The Scintillator Detectors}
\label{ssec1}
Complementing the water-Cherenkov detectors of the surface array with scintillator detectors will provide a way to
sample shower particles with detectors having very different responses to the muonic and electromagnetic 
components. 
Indeed, the ratio of the electromagnetic to muonic components of the showers is more than a factor of two higher in
a thin scintillator with respect to  the WCD over a very wide range of distances to the core.\\
A Surface Scintillator Detector (SSD) unit \cite{PrimeSSD} is composed of a box of (3.8 $\times$ 1.3) m$^2$, housing 
two scintillator sub-modules, each covering an area of 1.9 m$^2$. The scintillator planes are protected 
by light-tight, waterproof enclosures and are mounted on top of the existing WCD with a strong support frame, 
as shown in the left panel of Fig.\ref{fig_1}. 
A double roof, with the upper layer being corrugated aluminum, is used to reduce the temperature variations.
The scintillation light is collected with wavelength-shifting fibers 
inserted into straight extruded holes in the scintillator planes (right panels of Fig.\ref{fig_1}). 
The fibers, Kuraray Y11(300)M S-type, are positioned following 
the grooves of the routers at both ends, in a ``U'' configuration that maximises  the light yield. They are 
bundled in a cookie, acting also as a diffuser, whose PMMA front window is connected to a single photomultiplier 
(PMT) tube. The PMT is a bi-alkali  Hamamatsu R9420, 1.5" diameter,
with $\sim$18\% quantum efficiency at a wavelength of 500 nm; the power supply is based  on a custom made design 
manufactured by ISEG company.\\
After production, the SSD modules are tested using  muon tower facilities, exploiting atmospheric muon tomography 
to measure the  minimum ionizing particle (MIP) as well as the uniformity 
for each SSD unit. From these measurements, 
the number of photoelectrons per vertical MIP is found to be (30$\pm$2). Due to the U-turn of the fibers and their
length outside the scintillator bars, the uniformity of the measured signals is $\pm$5\% along the bars and $\pm$10\% 
between bars. The attenuation length of the light in the fibers is $\lambda=(312\pm3)$ cm.\\
Each SSD module is equipped with two sensors to measure temperature and humidity; by measurements in the field,
it has been shown that the temperature inside the SSD closely follows the air temperature evolution measured in 
the shade, almost never exceeding 40$^{\circ}$C.

%--------------------------------------------------------  SPMT
\subsection{The dynamic range of the Surface Detector}
The SD data quality will be improved by extending the acquisition dynamic range in both detectors, thus 
allowing us to measure non-saturated signals at distances as close as 250 m from the shower core \cite{PrimeSPMT}. 
In order to benefit the most from the combined information of SSD and WCD, the dynamic 
range of both detectors should be similar. Significant signals are expected both close to the shower core and  
at intermediate distances, where simultaneous measurements will be most important for the separation 
of the different components of the shower, furthermore allowing a direct cross-check of the two detectors.

To  increase the dynamic range, a fourth small PMT (Hamamatsu R8619, 1" diameter) is 
inserted in the WCD, exploiting an hitherto unused and easily accessible 30 mm window on the Tyvek bag containing
the ultra-pure water. With an active area of about 1/80 with respect to the large WCD PMTs, it potentially allows 
for an equivalent dynamic range
extension. The required range up to about 20000 VEM can be obtained by adjusting the gain in such a way that the
ratio of the large to the small PMT signals be limited to a value of 32.\\
The linearity of the small PMT is guaranteed in the required range for a large set of voltages, down to 800 V.
The power supply  is provided by an external custom made HVPS manufactured by CAEN company.

For consistency with the associated WCD, the dynamic range in the SSD must span from the signal of a single particle, 
as needed for calibration, to large signals, up to $\sim 2\times 10^4$ MIP.  Indeed, the SSD PMT has
been chosen also based on its excellent linear response when operated at low gain, being linear
within 5\% for peak currents up to 160 mA (for a gain of $8\times10^4$).  
To reach the required dynamic range, the 
anode signal is filtered and split in two in the electronics front-end; then one of the two signals is attenuated 
by a factor of 4, while the other is amplified by a factor of 32.
%%%%%%%%%% fig2
	\begin{figure*}
	\centering
	\includegraphics[width=0.85\linewidth,clip]{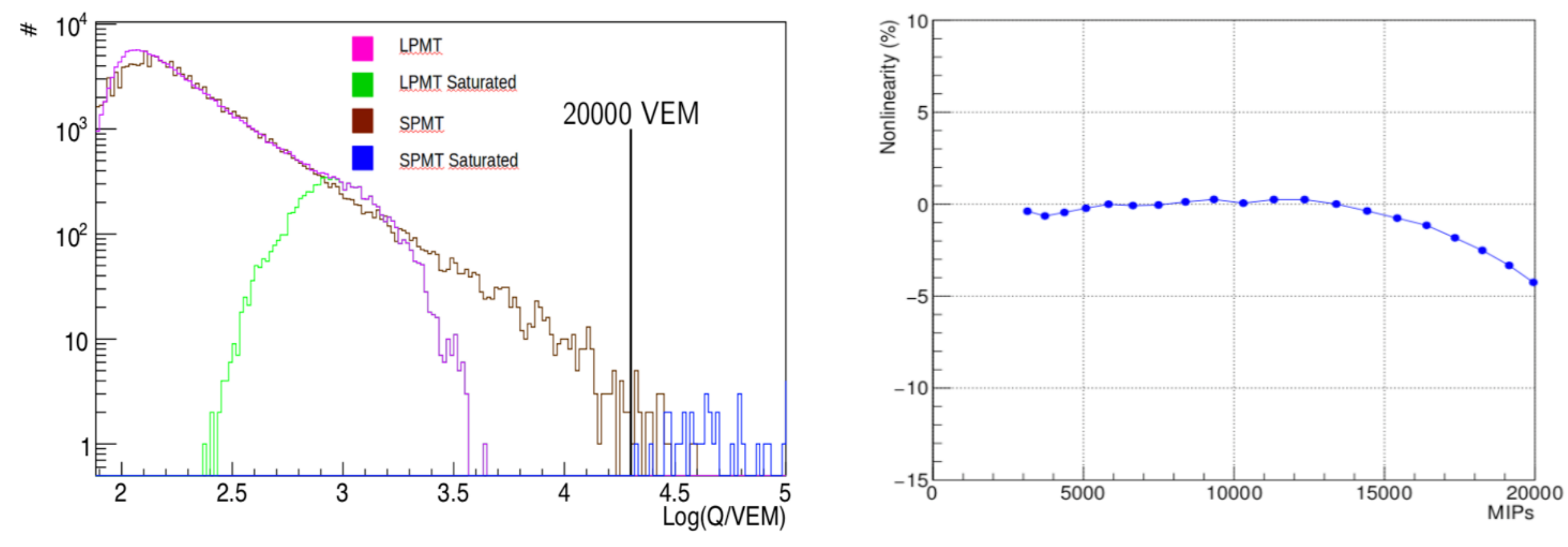}
	\vspace{-0.3cm} 
	\caption{{\it Left:} Charge spectrum in one of the upgraded WCD stations, as measured by the large PMTs 
	(pink: unsaturated, green:saturated) and by the small PMT (brown: unsaturated, blue=saturated). 
	{\it Right:} Linearity behaviour of one of the SSD-PMT.}
	\label{fig_2}   
	\end{figure*}
%%%%%%%%% fig3
	\begin{figure*}
	\centering
	\includegraphics[width=0.85\linewidth,clip]{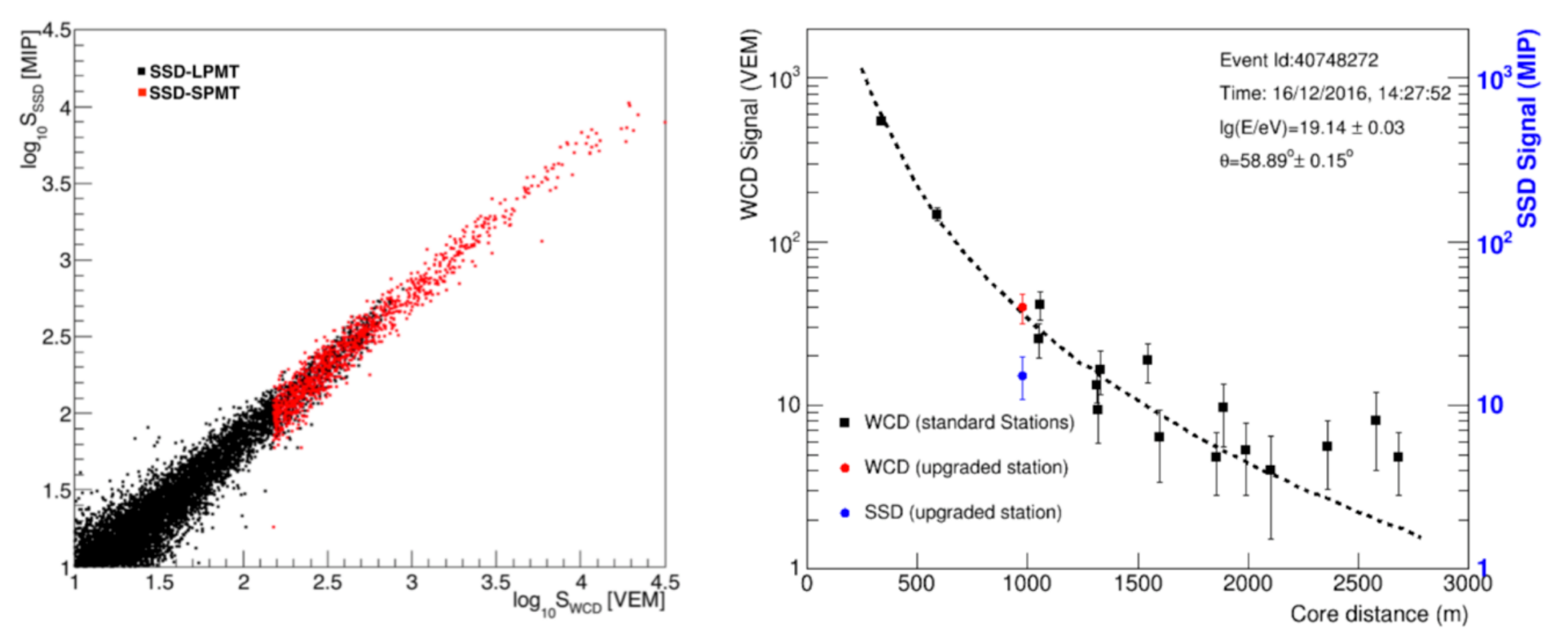}
	\vspace{-0.3cm} 
	\caption{{\it Left:} Correlation of signals in SSD vs WCD. The large PMTs are used up to the saturation (black dots); 
	the measurements are extended further by means of the small PMT (red dots). {\it Right:} Lateral distribution of an event close to the Engineering Array}
	\label{fig_3}   
	\end{figure*}

%-------------------------------------------------------- SDEU 
\subsection{The Electronics of the Surface Detector}
Implemented in a single board, called the upgraded unified board (UUB), the electronics  of the SD is used to record 
the PMT signals, make local triggering decisions, send time-stamps to the central data acquisition system 
and store event data for retrieval upon a global trigger condition \cite{PrimeSDEU}.  
The upgraded electronics will allow us to read the four photomultipliers of the WCD and the SSD one, processing the 
signals at 120 MHz, corresponding to an increase of the current rate of  a factor of 3. \\
The anode channel of the three large XP1805 PMTs of the WCD is split and amplified to reach a gain ratio of 32.
The signals are filtered and digitized by commercial 12 bit 120 MHz flash ADC (FADC); as the XP1805 pulse
response in terms of bandwidth is $\sim$70 MHz, this is well matched to a 120 MHz FADC and associated 60 MHz Nyquist filter.\\
The  small PMT signal is also digitized with 12 bits at 120 MHz in a separate channel. \\
The anode of the SSD PMT is split in two different channels, the first one amplified to have a gain ratio of 32 and 
the other attenuated by a factor of 4. This yields a total gain ratio of 128. The signals are filtered and digitized 
similarly to the WCD PMT signals. \\
The local trigger and processing capabilities are increased by using a more powerful local-station processor and 
FPGA (field-programmable gate array), thus allowing the implementation of new triggers, e.g. combined SSD and WCD 
triggers, in addition to the current ones. \\
The recorded traces can be digitally filtered and downsampled to 
40 MHz to emulate the current triggers in addition to any new ones, thus providing backwards-compatibility 
with the current dataset. Furthermore, this will allow us to deploy the new electronics without stopping or 
disturbing the current data taking.\\
The timing accuracy is improved to 2 nanoseconds thanks to the implementation of a new GPS receiver, the 
I-Lotus M12M, providing the same pulse per second timing output with serial control and data as the currently
used Motorola Oncore UT+ one, as such providing  compatibility with the running electronics. \\
The UUB is equipped with a micro-controller (MSP430) for the control and monitoring of the PMT high voltage, 
the supervision of the various supply voltages and reset functionality. More than 90 monitoring variables  
will be managed by the slow-control software.

The VEM (vertical equivalent muon) is the reference calibration unit  of the WCD high gain channels. It is derived
using a test station with an external hodoscope triggering on atmospheric muons and corresponds on average
to 95 photoelectrons at the PMT cathode; in the UUB, this is equivalent to about 1800 integrated ADC counts above pedestal. \\
Due to its dimensions, no direct calibration of the small PMT with muons is feasible. By selecting small shower events, 
the small PMT can be cross-calibrated with the large PMTs in such a way that the signal spectrum is correctly reconstructed 
up to about 20,000 VEM.  \\
The SSD calibration is based on the signal of a minimum-ionising particle going through the detector.  
A cross-trigger with the WCD can be used to remove all of the low-energy background. About 40\% of the 
calibration triggers of the WCD produce a MIP in the SSD. \\
For both the WCD and the SSD,  the cross-calibration between high gain and low gain channels is defined in the electronics, 
by a simple ratio between the amplified and non amplified anode signals.

%-------------------------------------------------------- Performances of the upgraded detectors
\subsection{Performances of the upgraded detectors}
Started in 2016, an Engineering Array of twelve AugerPrime stations has been taking data in the field, 
with the aim of testing and monitoring the performances of the upgraded detectors.

The logarithm of the charge spectrum in one of the upgraded AugerPrime WCD stations is shown in Fig.~\ref{fig_2}(left).
The dynamic range is extended to more than 20000 VEM by means of the small PMT, as one can see by 
comparing the unsaturated spectrum measured by the large PMTs (magenta histogram) to the one obtained with
the small PMT (brown curve).  The reconstruction resolution obtained with the small PMT is better than 5\% 
above 3000 VEM, which correspond to the signal produced at 250 m from the core in showers of 10 EeV.\\
The very good linearity behaviour of the photomultipliers of the SSD guarantees the possibility to measure
signals up to about 20000 MIP. As shown in the right panel of the same figure for one SSD-PMT, the non-linearity
stays below 5\% up to the highest densities.\\
The correlation between the signals in one of the WCD and in the corresponding 
scintillator is shown in the left panel of Fig.~\ref{fig_3}, where both scales are expressed in the corresponding physical units 
(VEM for the WCD PMTs and MIP for  the SSD PMT). The signals in the WCD are measured  up to saturation 
($\sim$ 650 VEM) by the large PMTs (black dots). In the superposition region and above the large PMT saturation,
they are derived from the small PMT (red dots). The correlation obtained demonstrates the validity of the
independent calibration procedures and shows that the required dynamic range in the upgraded Surface Detector is  
nicely covered up to the highest particle densities.  \\
The reconstructed lateral distribution function for one of the events collected in the region of the Engineering Array
is shown in the right panel of Fig.~\ref{fig_3}.
The signal recorded in the upgraded WCD (red dot) is in very good agreement with expectations, following
the lateral distribution reconstructed by means of the standard stations. The SSD signal, in blue, is relatively
lower than the WCD ones, as expected due to the different sensitivity of the two detectors to the electromagnetic 
component of the showers at different core distances and to the relatively smaller area of the SSD with respect to the WCD. 

%-------------------------------------------------------- AMIGA
\subsection{The Underground Detector}
\begin{figure}[h]
	\centering
	\includegraphics[width=0.75\linewidth,clip]{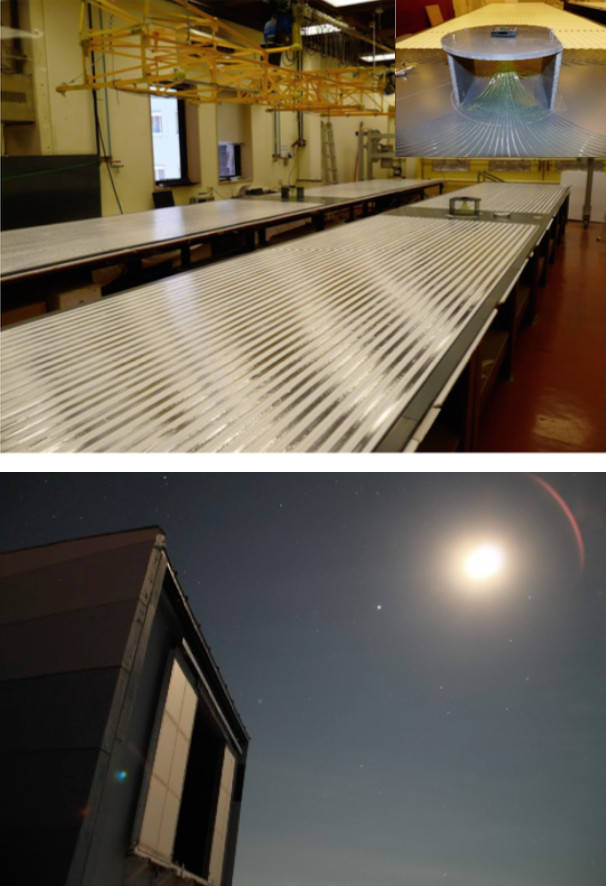}
	\vspace{-0.3cm} 
	\caption{{\it Top:} Two  AMIGA modules during assembly. {\it Bottom:} the Moon and Jupiter
	above one of the Fluorescence Detectors in the Pierre Auger Observatory.}
	\label{fig_4}   
	\end{figure}

As part of the AMIGA enhancement (Auger Muons and Infill for the Ground Array), an array of 
underground detectors  is being deployed beside the 61 WCD of the Infill area of the Observatory, on a 750 m triangular grid.  
In each position, 3 scintillator modules of 10 m$^2$ area are buried 2.3 m underground, 
to minimise the contamination from electromagnetic shower particles. The scintillation light is collected
using wavelength shifting fibers read by silicon photo sensors \cite{AMIGA}. Two modules in the
assembly hall are shown in the top panel of Fig.\ref{fig_4}.\\
The segmented structure of the counters allows direct counting of muons; a channel for signal integration will be used
to extract the muon information close to the shower core, in the saturation region.\\
These detectors will provide a direct measure of the muon content of the extensive air showers, with the aim of
studying the primary composition and hadronic interactions in a region of energy corresponding to the
transition of cosmic rays from a Galactic to an extra-galactic origin. \\
Furthermore, the underground modules will allow us to cross-check  and  fine-tune the methods used 
to derive the muon information from the upgraded SD.
%%%%%%%%% fig_4

%-------------------------------------------------------- FD
\subsection{The extended duty cycle of the Fluorescence Detector}
\label{ssec4}
The Auger Fluorescence Detector (FD) is composed of 27 telescopes overlooking the SD area (Fig.\ref{fig_4}, 
bottom panel). They are used to
observe the longitudinal development of air showers in the atmosphere, thus providing mass sensitive observables and
a model independent energy reconstruction.  \\
The FD duty cycle is currently  limited to about 15 \%, being operations 
limited to dark and moonless nights.
The current criteria for FD measurements require the Sun to be more than 18$^\circ$ below the horizon, the 
Moon to be below the horizon for more than 3 hours and the illuminated fraction of the Moon to be below 70\%.\\
A significant increase of about 50\% in the duty cycle can be obtained by extending the FD operations to times 
with larger night sky background, relaxing the second and third requirements. 
By reducing the supplied high voltage,  the PMT gains can indeed be reduced by a factor of 10. \\
Preliminary tests
show that the PMTs operated at reduced gain satisfy the criteria required for the FD performance (such as linearity, 
stability and lifetime), avoiding too high anode currents that could result in a deterioration of the tubes. 

%-------------------------------------------------------- Radio
\subsection{The Radio Upgrade}
\label{ssec5}
\begin{figure}[h]
	\centering
	\includegraphics[width=0.7\linewidth,clip]{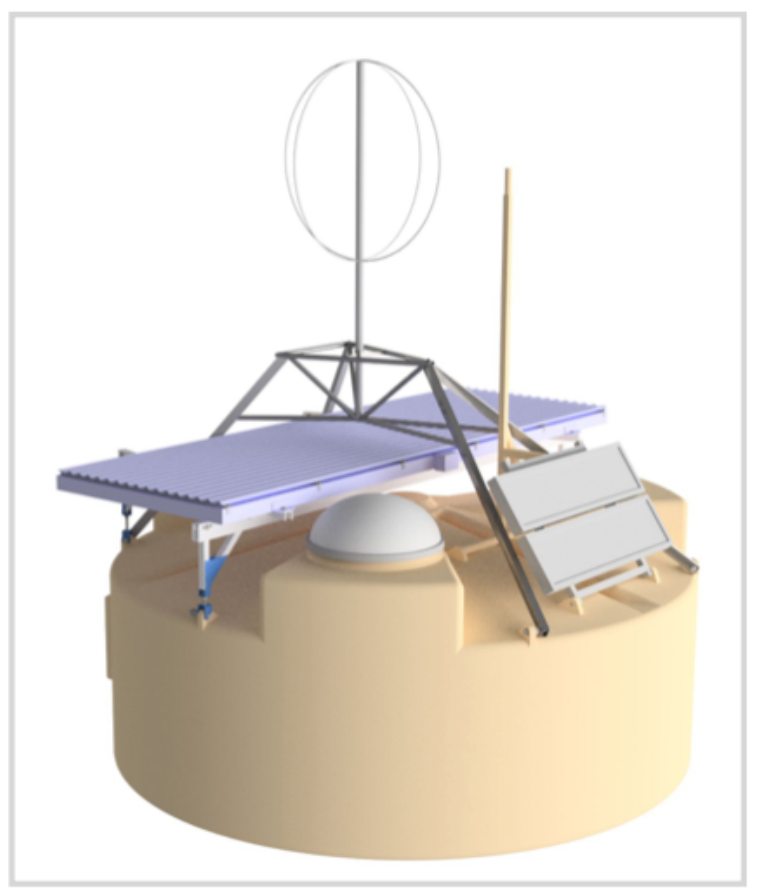}
	\vspace{-0.3cm} 
	\caption{Sketch of the radio antenna mounted atop the WCD+SSD.}
	\label{fig_5}   
	\end{figure}

An engineering array of radio detectors  (AERA) has been running in the Pierre Auger Observatory for a few years.
With the collected data, it was possible to demonstrate the feasibility of radio detection with a grid 
of antennas at 1500 m mutual distances, and measure horizontal showers ($\theta=60^{\circ}-84^{\circ}$), which illuminate an area of 
several km$^2$ on the ground  \cite{radioHor}.\\
Based on these results, a full radio upgrade of the Pierre Auger Observatory has been proposed  \cite{radio}, 
where each WCD of the SD will be equipped with a radio antenna mounted on its top surface, as shown in Fig.\ref{fig_5}.\\
The new detectors will operate together with the upgraded SD, forming a unique 
setup to measure the properties of cosmic rays above $10^{17.5}$ eV.  For horizontal showers, they
will nicely complement the information on the particle type  we will get from the WCD+SSD, thereby 
increasing the exposure for mass-sensitive investigations. 
%%%%%%%%% fig_5
%%%%%% fig_6
	\begin{figure*}
	\centering
	\sidecaption
	\includegraphics[width=0.9\linewidth,clip]{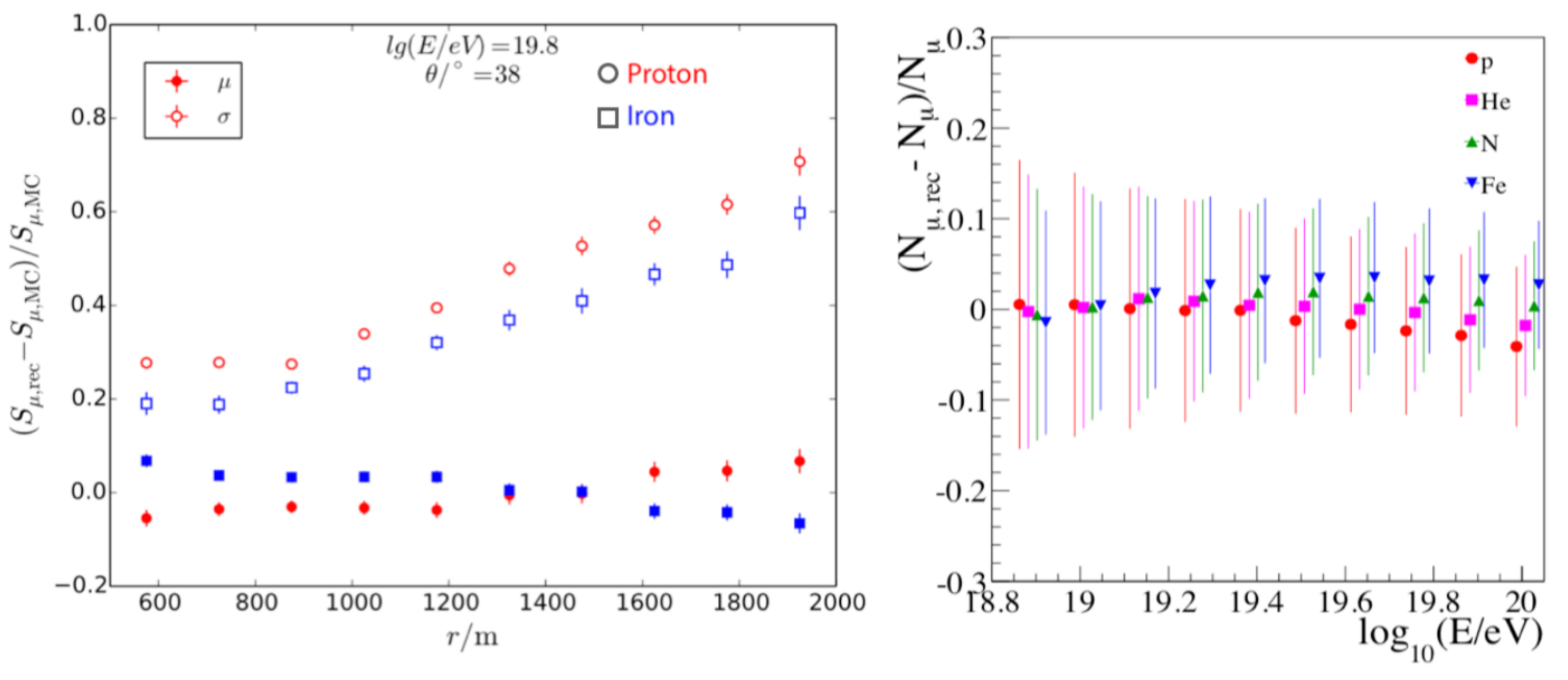}
	\caption{{\it Left:} Reconstruction bias (full symbols) and resolution (empty ones) for the muonic 
	signal derived with the matrix inversion method in individual stations, as a function of distance to the 
	shower core, for protons (circles) and iron nuclei (squares). {\it Right:} Relative difference
	between the number of muons reconstructed in individual showers using shower universality and the true (Monte Carlo) ones 
	as a function of energy for different primary species. Error bars represent the RMS of the distributions}
	\label{fig_6}
	\end{figure*}

%-------------------------------------------- Physics performances
\section{Expected physics performances}
\label{sec3}
As discussed above, the main aim of AugerPrime is that of providing event-by-event composition-sensitive 
measurements by increasing the ability to recognize and separate the different components of air showers. 
The goodness of the separation between two different primaries  {\it i} and {\it j} can be quantified by defining
a merit factor 
\begin{equation}
f_{MF}=\frac{|S_{i}-S_{j}|}{\sqrt{\sigma^{2}(S_{i})+\sigma^{2}(S_{j})}}
\end{equation}

The most direct technique to evaluate the muon component exploits the different responses of the WCD and 
SSD at station level, relating the total signals $S_{WCD}$(VEM) and $S_{SSD}$(MIP) to the electromagnetic 
and muonic fluxes at ground. 
Employing a matrix inversion method \cite{ALS}, the muonic signal in the WCD can be derived as 
\begin{equation}
S_{\mu,WCD}={\it a} S_{WCD} + {\it b} S_{SSD}
\end{equation}
where the coefficients {\it a} and {\it b}, although derived from simulations, show only a very weak dependence on
the mass and arrival direction of the primary particles and on the hadronic interaction models used in the Monte Carlo.\\
The muon signals so obtained are subject to large fluctuations, being restricted to individual stations.
%%%%%% fig_7
	\begin{figure*}
	\centering
	\sidecaption
	\includegraphics[width=0.99\linewidth,clip]{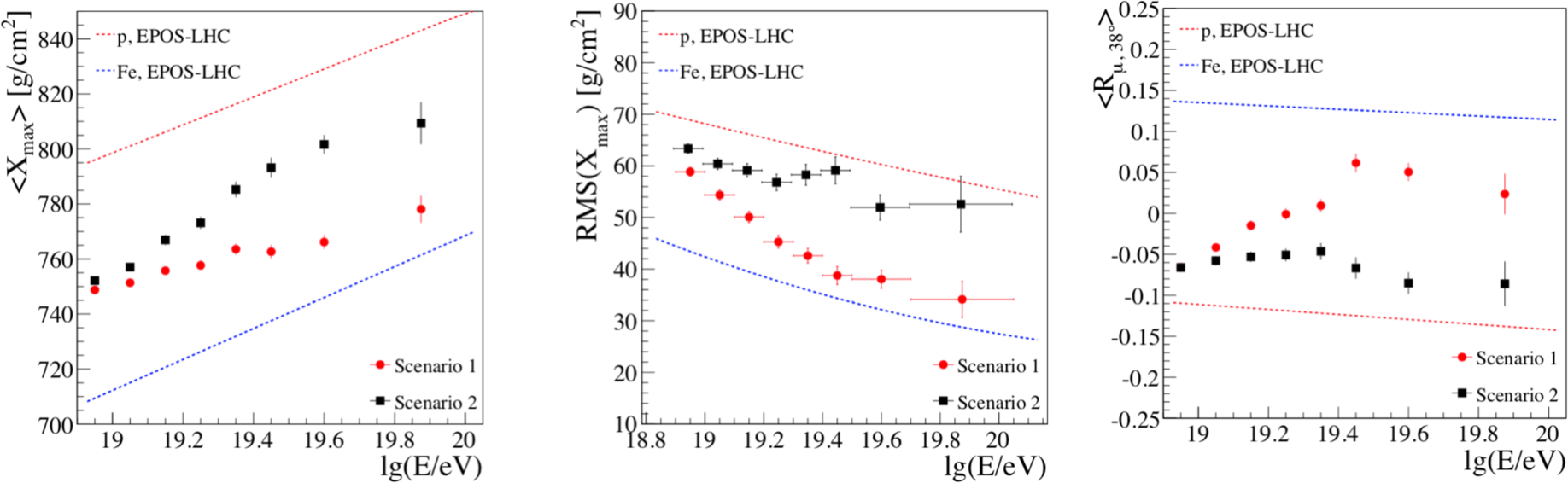}
	\caption{Mean depth of shower maximum $X_{max}$ (left), its fluctuations (center) and the relative muon number
	at $38^\circ$ (right), evaluated for the two scenarios (1=maximum rigidity model; 2=photo-disintegration model) using 
	simulated SD data only, for pure proton and iron compositions.}
	\label{fig_7} 
	\end{figure*}
In Fig.\ref{fig_6}, left,  the reconstruction bias (solid symbols) and the resolution (open symbols) of the muonic signal 
contribution for individual detector stations are shown for two different primary compositions: at 1000 m from the core, a
resolution of 20 to 30\% can be achieved.\\
The performances of the method can be improved when the reconstructed signals at station level are used to evaluate 
the lateral distribution function. The matrix inversion method is then applied to evaluate the muon signal at 
800 m from the core. Using $S_{\mu}(800)$ as estimator of the composition, a merit factor $f_{MF}>1.5$ can be 
obtained in separating proton and iron primaries at small zenith angles for energies larger than $10^{19.5}$ eV.

A more general approach based on the properties of universality of air showers \cite{univ} or multivariate analyses would allow us 
to take advantage of all the information provided by AugerPrime, not only by correlating the detector signals at 
different distances to the core but also considering their temporal structure (arrival times of the particles at the stations). 
In this case, an event-by-event reconstruction allows us to derive not only the energy, but also the 
depth of maximum shower development in 
atmosphere ($X_{max}$) and the number of muons ($N_{\mu}$) using  the information from the scintillators
combined with the WCD data.
The bias and resolution in the evaluation of $N_{\mu}$ are shown in the right panel of Fig.\ref{fig_6}, 
as a function of composition and energy. As an example, a merit factor of about 1.7 (2.1) can be obtained for the
separation of proton from iron primaries at 60 EeV, for zenith angles of 38$^\circ$ (56$^\circ$) respectively. 
The reconstruction bias for $X_{max}$ is less than 15 g/cm$^2$ with the resolution improving from 40 g/cm$^2$ at 
$10^{19}$ eV to 25 g/cm$^2$ at $10^{20}$ eV.

\subsection{The interpretation of the suppression in the flux}
To explore the power of AugerPrime with respect to meeting the scientific goals, we simulated two 
sets of mock data corresponding to seven years of data taking in AugerPrime and to an exposure of 
50000 km$^2$ sr yr, assuming the two different simplified models described 
in Sect.\ref{intro}: a maximum rigidity (``Scenario 1'') and a photo-disintegration 
one (``Scenario 2''). These two benchmark models approximately reproduce the energy spectrum and  $X_{max}$
distributions as measured in Auger above $10^{18.7}$ eV, as resulting from a numerical fit 
following the method described in  \cite{scenarios}. \\
The first two moments of the $X_{max}$ distributions and the mean muon number as predicted based on the two
scenarios are shown in Fig.\ref{fig_7} as a function of energy for two extreme nuclear masses. Note that the RMS($X_{max}$)
includes both the shower-to-shower fluctuations and the detector resolution.\\
To get rid of the dependencies of the total muon number on energy and zenith angle (due to the shower intrinsic fluctuations 
and to the fact that the muon path-length varies with $\theta$), the total muon number is shown relative to that 
expected from a mix of equal fractions of primaries.  
 Although the expected values are quite similar below $10^{19.2}$ eV,
a range already covered by the FD, the models predict quite different results in the highest energy region, showing 
that the two scenarios can be distinguished with high statistics and significance in 7 years of measurements. 

\subsection{Is there a proton fraction at UHE?}
\begin{figure}[h]
	\centering
	\includegraphics[width=0.8\linewidth,clip]{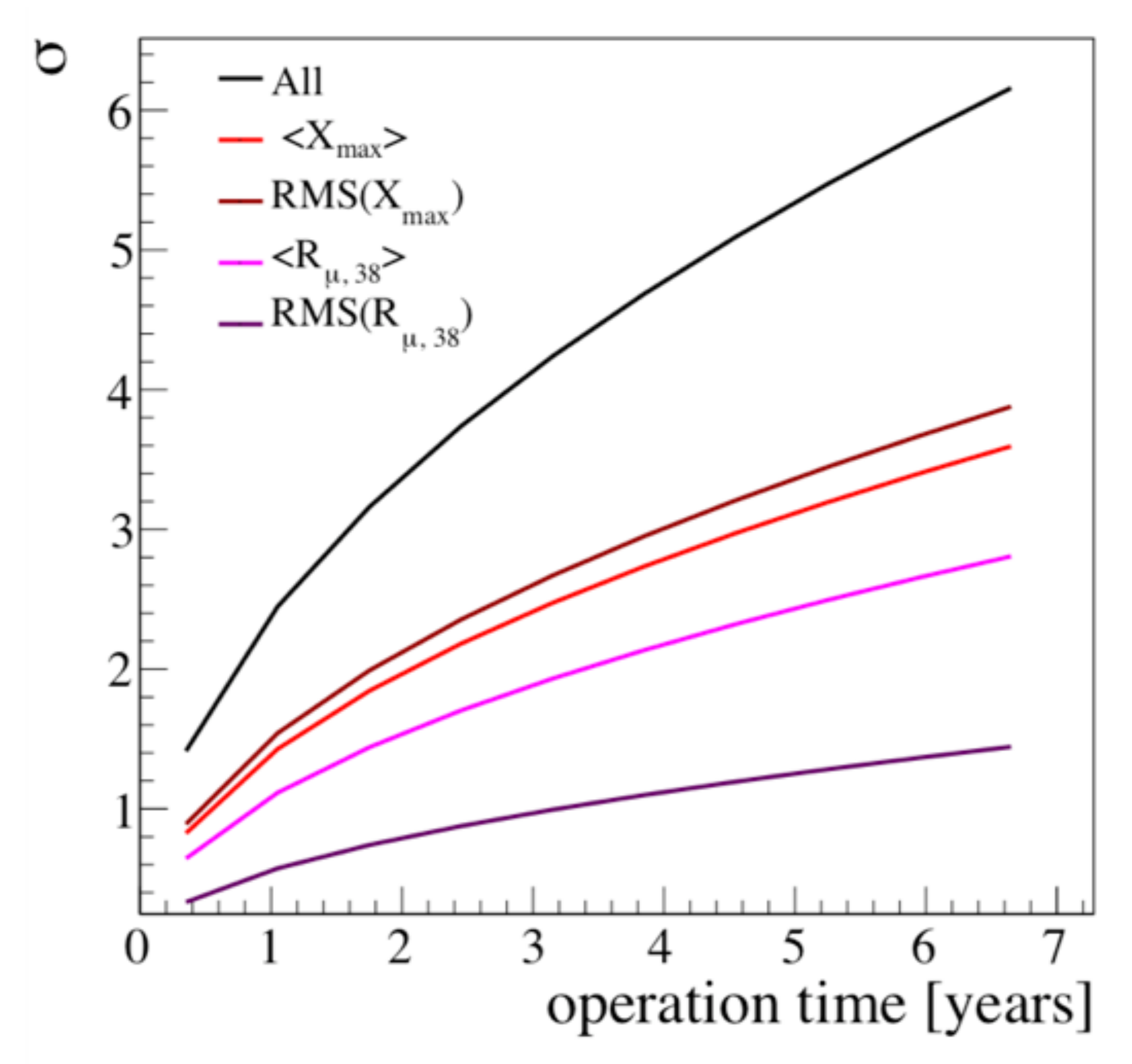}
	\caption{Significance of distinguishing Scenarios 1p and 1 (with and without at least 10\% protons respectively) 
	as a function of the operation time of AugerPrime.}
	\label{fig_8}  
	\end{figure}

In the first scenario, no protons can reach us from above 50 EeV, while on the contrary we would see a large fraction
of them in the second one. \\
To explore the feasibility of detection of a small fraction of protons with AugerPrime in the
framework of Scenario 1, we artificially added a contribution of 10\% protons to it (``Scenario 1p''). 
The proton-rich sample was produced by assigning to these events a proton-like $X_{max}$, i.e. an $X_{max}>700$ g/cm$^2$
at 10 EeV, adjusted to the event energy with an elongation rate of 55 g/cm$^2$ per decade.\\
The difference $\sigma$ between the means of the distributions as obtained from the two scenarios (``1'' and ``1p'')
divided by the statistical uncertainty is shown in Fig.~\ref{fig_8} for each observable. As shown in the figure by the black line, 
the best separation power is obtained by constructing a combined significance using all
observables at our disposal:
\begin{align}
\sigma^{2} &= \sigma^{2}(<X_{max}>)+\sigma^{2}(RMS(X_{max})) \nonumber \\
                  &+ \sigma^{2}(<R_{\mu,38^{\circ}}>)+\sigma^{2}(RMS(R_{\mu,38^{\circ}}))
\end{align}
We conclude that using this combination and integrating above 10 EeV we might be 
able to distinguish between Scenario 1 and 1p (i.e. to observe a fraction of protons as low as 10\%) at the highest 
energies with more than 5$\sigma$ after 5 years of operation of  AugerPrime.
%%%%%% fig_8
%%%%%% fig_9
	\begin{figure*}
	\centering
	\includegraphics[width=0.99\linewidth,clip]{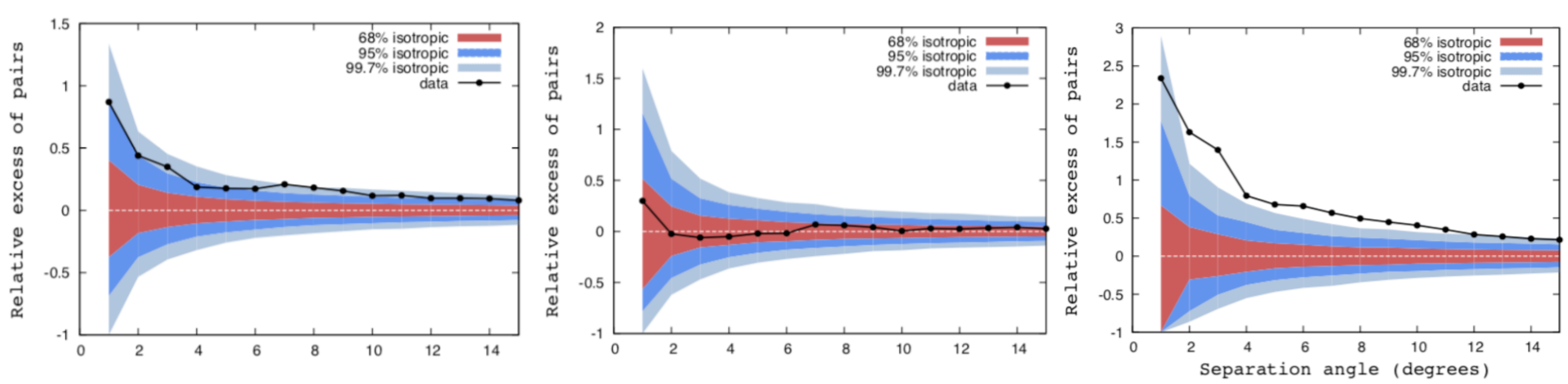}
	\caption{Angular correlation of Auger data modified according to Scenario ``1'' and ``1p'' with the AGNs of the ``70 Months Swift-BAT
	catalog'' (see text for details). The relative excesses of pairs of events as a function of their angular separation is shown for
	the complete data set (left), the selection deprived of light elements (center) and the proton-enriched one (right panel).}
	\label{fig_9}   
	\end{figure*}
%%%%%% fig_10
	\begin{figure*}
	\centering
	\includegraphics[width=0.80\linewidth,clip]{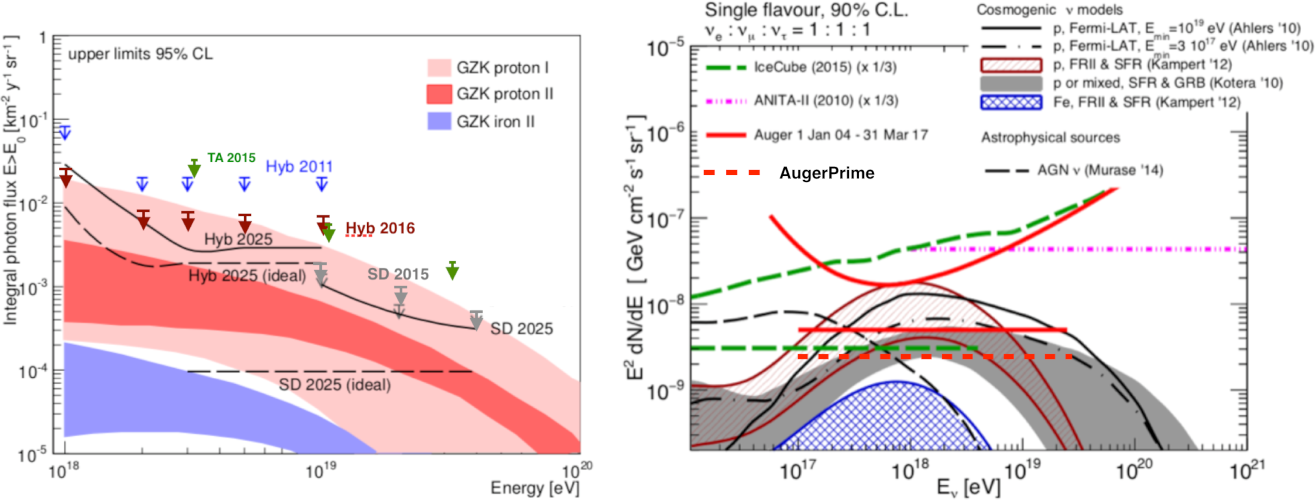}
	\caption{Expected sensitivity on the flux of photons (left) and neutrinos (right) \cite{phot,neut}. For photons, full black lines show 
	more conservative estimates based only on the increase of statistics.}
	\label{fig_10}   
	\end{figure*}

\subsection{Composition-wise anisotropy studies}
To study the gain in sensitivity with AugerPrime in the search for the sources of ultra-high energy cosmic
rays, we used the arrival directions of 454 events with E $>4\times 10^{19}$ eV measured at $\theta < 60^\circ$ \cite{ev454}, 
assigning them random $X_{max}$ values according to Scenario 1. A 10\% enhanced proton fraction was then 
implemented as in the previous study (Scenario 1p). Furthermore, we randomly assigned to 50\% of these
protons an arrival direction corresponding within 3$^\circ$ to one of the AGNs of the Swift-BAT catalog, 
chosen among those with distances within 100 Mpc \cite{Swbat}.\\
The angular correlation is shown in Fig.~\ref{fig_9}. It is not significant when
considering the complete data set of 454 events without composition information (left panel of the figure), but 
with the proton-enriched sample a correlation in excess of 3$\sigma$ is clearly evident (right panel).
As expected, no excess of pairs is found with the AGNs for the proton-deprived selection (central plot), as in
this case the angular correlation is just due to statistical fluctuations.

\subsection{Neutral particles limits}
Auger observations provided the currently most stringent limits on the flux of photons at energies above 
$10^{19}$ eV and severe constraints on  cosmogenic neutrinos \cite{phot,neut}.\\
Even better limits are expected to be reached with AugerPrime: five more years of data will provide us with a large
increase in statistics, the new electronics will allow the introduction of more efficient triggers, a reduction of the 
hadronic background will be possible thanks to an improved muon discrimination. \\
In Fig.\ref{fig_10}, the current limits for both photons and neutrinos are shown together with
preliminary calculations of the expected sensitivities from AugerPrime (dashed lines). This ideal
case assumes complete rejection of the hadronic background.  
In the photon case, the introduction of the new triggers allows us to extend the measurement with SD below 10 EeV.

\subsection{Hadronic interactions and fundamental physics studies}
\begin{figure}
\centering
\includegraphics[width=0.98\linewidth,clip]{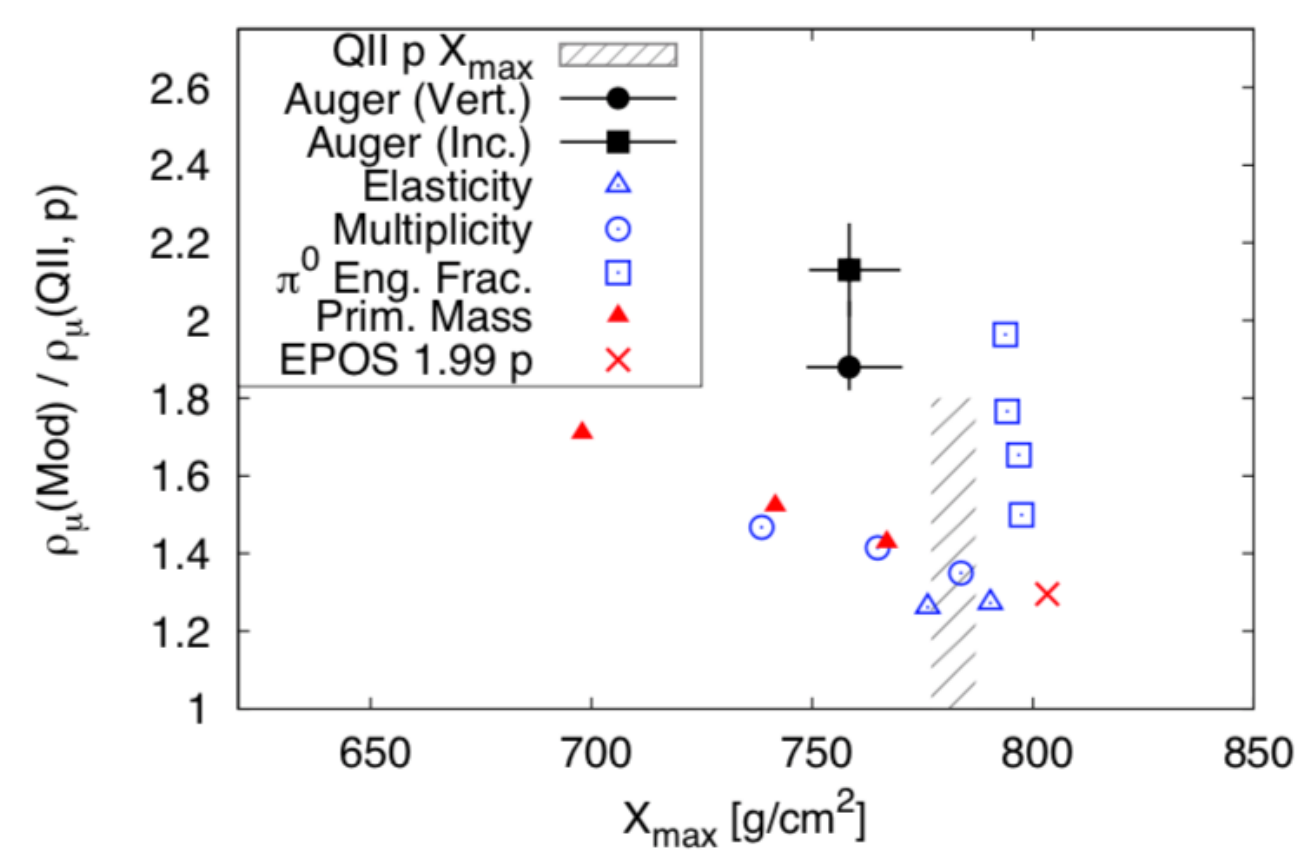}
\caption{Relative number of muons at 1000 m from the shower core wrt that predicted by QGSJETII-03 as a
function of $X_{max}$. Different modifications of hadronic interactions are considered  \cite{GF}. Auger data 
are also shown as derived from showers of $10^{19}$ eV for vertical and horizontal events.}
\label{fig_11}   
\end{figure}

AugerPrime can shed light on whether the inconsistencies between air shower simulations and 
experimental data are due to fundamental shortcomings in our understanding of hadronic multiparticle production.\\
Since the electromagnetic energy deposit in the atmosphere  is mainly due to the first high energy interactions,
while on the other hand muons at ground come as products of low energy interactions, 
an event-by-event correlation between the depth of shower maximum and the muon content in showers generated 
by ultra-high energy cosmic rays can bring strong constraints on hadronic interaction models \cite{GF}.
As an example, the muon density as obtained by different modifications of hadronic interaction models, 
relative to that predicted by QGSJetII-03 for protons, is shown in Fig.\ref{fig_11} in correlation with the depth of 
shower maximum, for various compositions.  

A high resolution measurement of the muon number such as the one expected from AugerPrime can prove to be 
a very powerful observable also to look for exotic interaction Scenarios.
As an example, if Lorentz invariance violation effects exist, the absorption and energy losses of UHECRs 
during propagation would be modified. Furthermore, strong changes in the development of shower produced by the interaction 
of UHECRs with nuclei in the atmosphere may also be visible \cite{exotic}.

%-------------------------------------------- Conclusion
\section{Conclusion}
\label{sec4}

The quest for the sources of UHECRs will be pursued by AugerPrime, the Pierre Auger Observatory upgrade, by
providing a high statistics sample of events with mass information.

Composition-wise anisotropy searches will allow us to understand the origin of the flux suppression and explore
the anisotropy dependencies on the particle rigidities.  AugerPrime will be able to measure the presence of even 
a small fraction of protons at the highest energies, as such assessing the feasibility of future projects willing to
detect ultra high energy neutrinos. 

A much improved measurement of the muon content in air showers will help in the study of hadronic 
multiparticle production and the search for exotic interactions. Furthermore, once a better tuning of the models is reached,
we will have the possibility to reanalyse also the pre-upgrade data set and study the composition with much lower
uncertainties.

While the main aim of the upgrade is the study of the highest energy region, above $10^{19.5}$ eV, the composition
sensitive information will also allow to gain more information on the region of transition from a Galactic to an extra-galactic 
contribution to the cosmic ray flux, at $E=10^{17}-10^{19}$ eV.

An Engineering Array of twelve upgraded stations has been taking data in the field since 2016; as of October 2018, 
thirty more fully upgraded stations have been deployed. The data collected so far demonstrate the quality of the
new detectors and the physics potential of the upgrade project. 
The production of  scintillators and small PMT systems is proceeding at full pace: a large number of them 
is already available in Malarg\"ue, while some units of the upgraded 
electronics boards are under test in the field. The production and deployment of the AugerPrime detectors and electronics
will be completed within 2019.
Operations and full data taking are planned from 2020 to 2025.
%%%%%% fig_11

\end{document}